\documentclass[11pt,a4paper]{article}

\usepackage[pdftex]{graphicx}
\usepackage{amsmath}
\usepackage{amssymb}
\usepackage{setspace}
\usepackage{subfigure}
\usepackage[usenames,dvipsnames]{color} 
\usepackage{marvosym}
\def\dotline{{\cdot\mkern-6mu\cdot\mkern-6mu\cdot\mkern-6mu\cdot\mkern-6mu\cdot}}

\begin{document}
\title{Intracellular Microrheology of Motile \emph{Amoeba proteus}}
\author{Salman S. Rogers, Thomas A. Waigh\thanks{thomas.waigh@manchester.ac.uk}, Jian R. Lu \\
Biological Physics Group, School of Physics and Astronomy, \\University of Manchester, Manchester M60~1QD, UK}
\maketitle

\section*{Abstract}
The motility of motile \emph{Amoeba proteus} was examined using the technique of passive particle tracking microrheology, with the aid of newly-developed particle tracking software, a fast digital camera and an optical microscope. We tracked large numbers of endogeneous particles in the amoebae, which displayed subdiffusive motion at short time scales, corresponding to thermal motion in a viscoelastic medium, and superdiffusive motion at long time scales due to the convection of the cytoplasm. Subdiffusive motion was characterised by a rheological scaling exponent of 3/4 in the cortex, indicative of the semiflexible dynamics of the actin fibres. We observed shear-thinning in the flowing endoplasm, where exponents increased with increasing flow rate; i.e. the endoplasm became more fluid-like. The rheology of the cortex is found to be isotropic, reflecting an isotropic actin gel. A clear difference was seen between cortical and endoplasmic layers in terms of both viscoelasticity and flow velocity, where the profile of the latter is close to a Poiseuille flow for a Newtonian fluid.

\section{Introduction}
Amoeboid locomotion is the general term describing the motility exhibited by adherent eukaryotic cells that move by extending pseudopodia, cytoplasmic streaming and changing their shape. Amoeboid locomotion is exhibited by many types of cell, including free-living amoebae and mammalian cells. In the latter, amoeboid locomotion is vital for the development of the embryo, the repair of wounds, and the action of the immune system. It is also responsible for the spread of tumors. Physical theories of amoeboid locomotion have developed since the mid-19th century, when the role of contractions of the cell cortex was recognised (reviewed in \cite{debruyn1947}). Eventually the fundamental chemistry of this contractility was understood in terms of the acto-myosin system and its interaction with the cell membrane and other cellular components \cite{bray2001,taylor1979}. 
In recent years, the motion and rheology of cells as well as reconstituted cytoskeletal systems have been an active research field in physics. New techniques have allowed measurements of the forces exerted and experienced by cells \cite{harris1980,bausch1998,dembo1996,panorchan2006}, as well as the forces and dynamics of individual cytoskeletal molecules \cite{howard2001}. Theoretical progress has been made on the rheology and force generation of the cytoskeleton \cite{liverpool2001,granek1997,howard2001,janmey2004}, and on integrated models of specific examples of amoeboid and amoeboid-like motion \cite{joanny2005,paluch2006jcb,dembo1989,grebecki1990,yoshida2006}.

\begin{figure}
\centering
\resizebox{9cm}{!}{\includegraphics{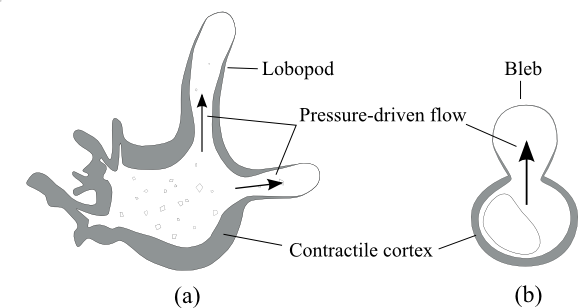}}
\caption{We have the same broad understanding of the mechanics of motion in a large amoeba (a) and cortical oscillations in a mammalian cell (b). Both are driven by contractions in the cortical cytoskeleton, causing pressure-driven flow into lobopodia or blebs. \emph{A.~proteus} is a useful model for microrheology, due to its thick cortical gel layer, and the large number of visible embedded particles.}
\label{corticalosc}
\end{figure}

Amoeboid locomotion has different characteristics in different cell types, and sometimes different characteristics in the same cell type in different environments \cite{taylor1979}. Large free-living amoebae such as \emph{Amoeba} and \emph{Chaos} exhibit lobopodia, with a clear distinction between the ectoplasmic gel layer and endoplasmic sol, which streams freely as the amoeba moves. Many mammalian cells on two-dimensional tissue-culture subtrates exhibit flat lamellipodia and thin filopodia, and comparatively slow locomotion. However, these cells often behave differently in three-dimensional tissue, where their extending pseudopodia may have closer resemblence to lobopodia. The similarity between motion in mammalian cells and amoebae has been further highlighted by observations of phagocyte motion, cortical oscillations in cells lacking microtubules \cite{paluch2005}, and cytokinesis \cite{paluch2006jcb,he1997dembo}. In both cortical oscillations of mammalian cells and crawling large amoebae, movement is generated by the contraction of the cytoskeleton in the cortical layer, which forces the cytoplasm into an expanding lobopod or bleb respectively (see Fig.~\ref{corticalosc}). 

We present a novel perspective on amoeboid locomotion using a new experimental technique---passive particle-tracking microrheology (PTM). In PTM, the thermal motion of tracer particles is tracked, yielding information on the rheology of the material \cite{waigh2005,levine2000,mason1997,gittes1998,Tseng2002}. A new advance is that we have developed software that allows endogeneous particles to be accurately tracked against a complex optical background i.e.~the cell. This software functions well with a fast digital camera (10KHz) and a high resolution oil immersion light microscope \cite{rogers2007}. Our experimental subject is \emph{Amoeba proteus}, which has been a popular model of amoeboid locomotion for over a century \cite{taylor1979}. For our technique, \emph{A.~proteus} is especially useful due to its large number of endogeneous particles which are visible in the light microscope, and its thick cortical gel layer.

\section{Methods}
\subsection{Amoeba culture}
Specimens of \emph{A.~proteus} were cultured by setting up a food chain according to \cite{page1976}. The culture medium consisted of 200 ml of Prescott's and James's solution with two boiled rice grains. Samples of \emph{A.~proteus} and the cryptomonad flagellate \emph{Chilomonas} were obtained from Blades Biological Ltd.~(UK) and added to the medium in a glass dish. \emph{A.~proteus} fed on \emph{Chilomonas} which fed on bacteria in the medium. The culture was kept at room temperature.

\subsection{Particle Tracking Microrheology}
Individuals of \emph{A.~proteus} were observed in an Olympus IX71 inverted microscope (Olympus UK Ltd., London EC1Y 0TX, UK) in bright field mode, with a 100x oil-immersion lens and a 1.6x magnifier. Images were captured using a Photron FastCam PCI CCD camera (Photron (Europe) Ltd., Bucks SL7 1NX, UK), at framerates up to 3000~Hz. Tracking of endogeneous particles was performed using our recently-developed particle tracking software \cite{rogers2007}, which is especially suited for tracking low-contrast particles against a complicated background. Several large and small individuals were examined, and measurements presented below are representative of all sampled amoebae in similar conditions.

\subsection{Confocal Laser Scanning Microscopy (CLSM)}
Confocal fluorescence images were taken with a Leica SP2 AOBS confocal microscope (Leica Microsystems GmbH., 35578 Wetzlar, Germany). TubulinTracker (taxol conjugated to Oregon Green 488, Molecular Probes, Invitrogen Ltd, Paisley, UK) was added directly to the culture medium containing live amoebae. Staining of F-actin was performed with Alexa Fluor 633 phalloidin (Molecular Probes), after fixing amoebae in 3.7\% formaldehyde in PBS for 10 min at room temperature, and permeablising with 0.1\% Triton in PBS for 3 min. 

\section{Results}

\begin{figure}
\centering
\subfigure[]{\resizebox{!}{6cm}{\includegraphics{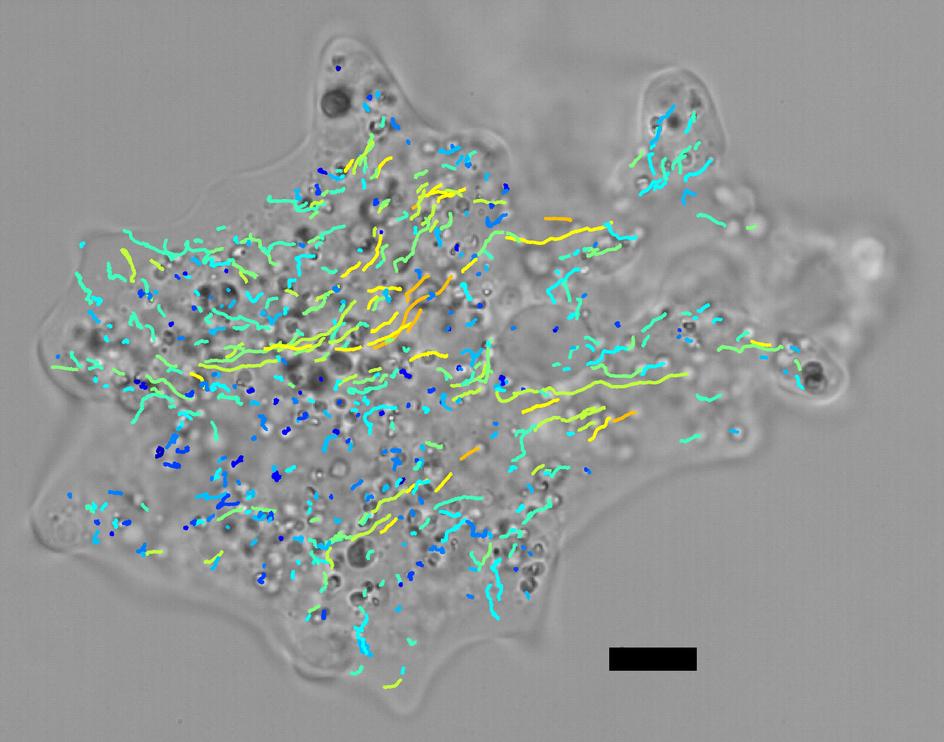}}}
\subfigure[]{\resizebox{!}{6cm}{\includegraphics{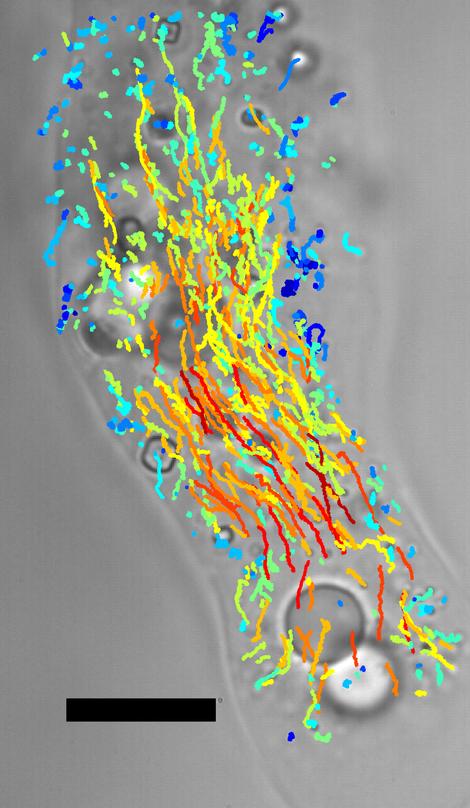}}
\resizebox{!}{6cm}{\includegraphics{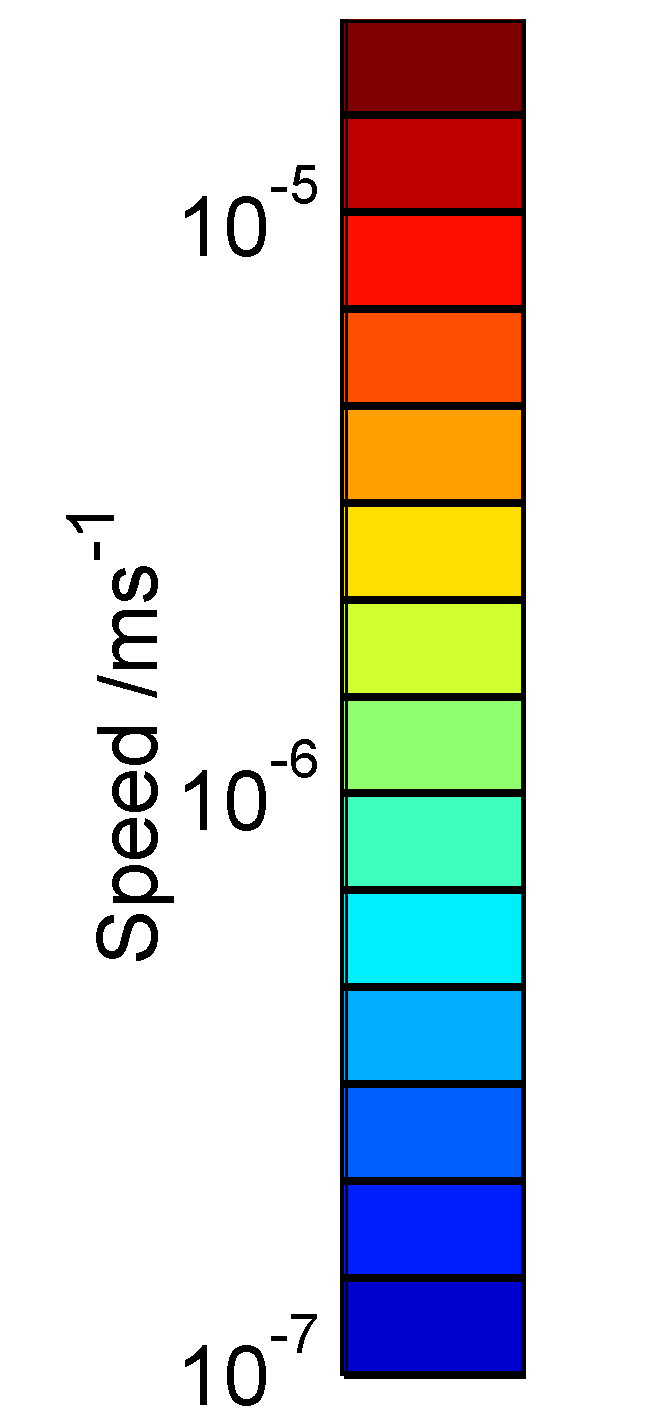}}}
\caption{(a) Small individual of \emph{A.~proteus} and (b) lobopod of large polytactic individual. 10~$\mu$m scale bars are plotted on each, and trajectories of all tracked particles are superposed, coloured according to their speed on the scale shown (right).}
\label{amoebae}
\end{figure}

Fig.~\ref{amoebae} shows (a) one small and (b) one large invididual of \emph{A.~proteus}. The large individuals are polytactic: they extend and contract lobopodia in several directions simultaneously. Fig.~\ref{amoebae}(b) shows a single pseudopod. Tracks of endogeneous particles are superposed on images (a) and (b), captured at 2000~Hz and 3000~Hz respectively. The average speed of each particle is represented by colours from blue (slowest) to red (fastest). It is clear that the flow of cytoplasm in the amoebae is inhomogeneous: the small amoeba shows localised fast currents surrounded by slow or gelled cytoplasm, while the large amoeba shows fast flow in a central channel (the endoplasm), surrounded by a much slower layer (the cortex or ectoplasm). These measurements reflect previous observations of intracellular velocities, e.g.~by \cite{kamiya1964}. 

\begin{figure}
\centering
\subfigure[]{\resizebox{8cm}{!}{\includegraphics{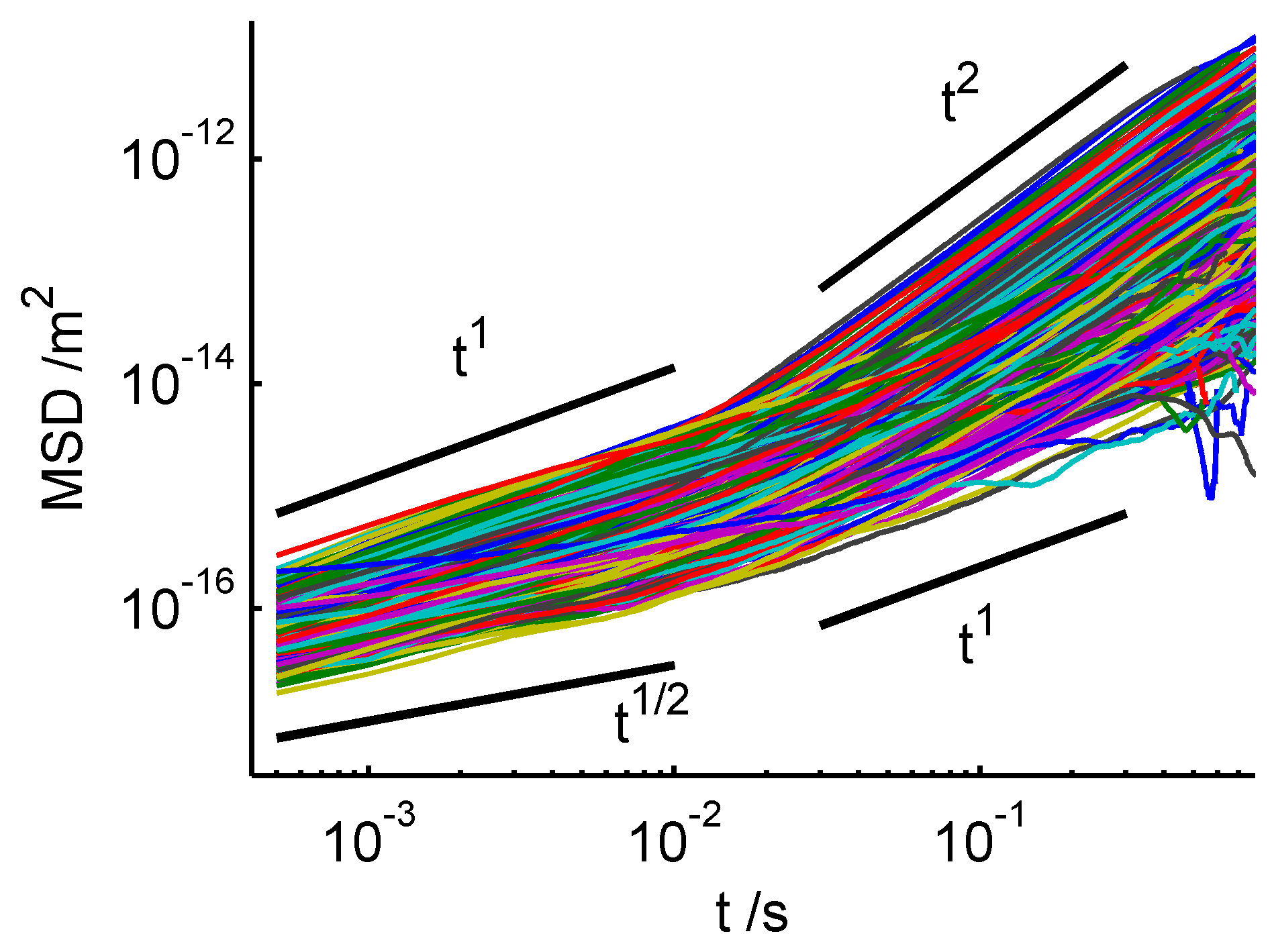}}}
\subfigure[]{\resizebox{8cm}{!}{\includegraphics{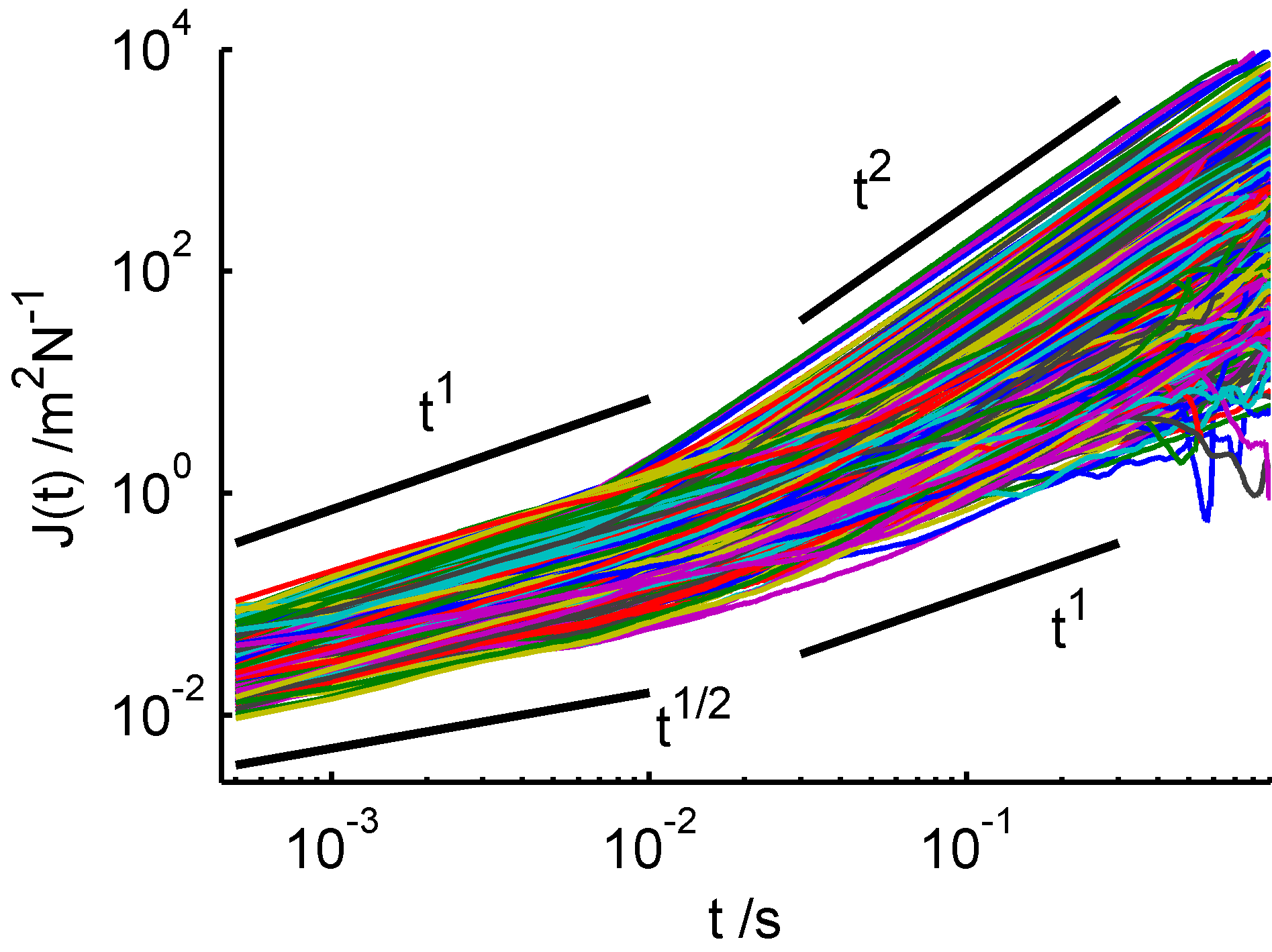}}}
\caption{(a) Mean-squared displacement (MSD) and (b) compliance $J(t)$ of all tracked particles in the small amoeba in Fig.~\ref{amoebae}(a), as a function of time. Two regimes are identified: a subdiffusive regime at small $t$, indicative of thermal motion in a viscoelastic medium, and a superdiffusive regime at large $t$, indicative of active motion.}
\label{smallaprot-MSD-J}
\end{figure}

From the particle tracks, we calculate the mean-squared displacement (MSD) of each track, as a function of time scale $t$, according to:

\begin{equation}
\textrm{MSD}(t)=\langle | \mathbf{r}(T+t)-\mathbf{r}(T) |^2 \rangle \,,
\end{equation}
where $\mathbf{r}(T)$ is the position at time $T$, and the average is performed over $T$. Fig.~\ref{smallaprot-MSD-J}(a) shows MSDs of all tracked particles in the small amoeba above. Two power-law regimes can be identified as different time scales. At large $t$, most particle tracks are superdiffusive---i.e.~dominated by active motion---since they have exponents greater than 1 and indeed as large as 2, which corresponds to steady, directed motion. At small $t$, motion is subdiffusive---corresponding to thermal motion in a viscoelastic medium---since the MSDs have exponents less than 1. The crossover takes place at $t \sim 0.02$~s, which reflects the rheology of the medium and the speed of the flow. The exponents at small $t$ may be understood according to the theoretical models of the rheology of stiff filaments: in particular, an exponent of 3/4 is expected for a network of tethered semiflexible filaments, and 1/2 is expected for semiflexible filaments under a sufficiently large tension \cite{gittes1998,granek1997,capsi1998}. 

As $t$ decreases to $0.5$~ms, the MSD curves decrease steadily with no apparent leveling-off due to the ``static'' error of measuring the particle positions \cite{Savin2005a}. Therefore our static error is insignificant compared to particle displacements, and must be significantly less than $\sqrt{\textrm{MSD(0.5~ms)}}=5$~nm.  

The MSD is related to the rheology of the material around each particle by the Generalised Stokes-Einstein Relation (GSER). We calculate the time-dependent shear compliance $J(t)$ which is directly related to the MSD \cite{xu1998}. In the case of a two-dimensional MSD:

\begin{equation}
J(\tau) = \frac{3 \pi a}{2 k_B T} \textrm{MSD}(\tau) \,,
\label{eq:compliance}
\end{equation}
where $a$ is the radius of each particle, which can be measured from the image (calculated according to \cite{rogers2007}). Fig.~\ref{smallaprot-MSD-J}(b) shows $J(t)$ for the same data as Fig.~\ref{smallaprot-MSD-J}(a). There is a large variation in absolute values of $J(t)$ between particles, even when the scaling exponent is the same. This variation is approximately a factor of 10 at $t=1$~ms. This is likely to be due, in part, to the local heterogeneity of the cellular environment \cite{Tseng2002} or the heterogeneity of contacts between the particle and the cytoskeleton \cite{Citters2006}.

\begin{figure}
\centering
\subfigure[]{\resizebox{8cm}{!}{\includegraphics{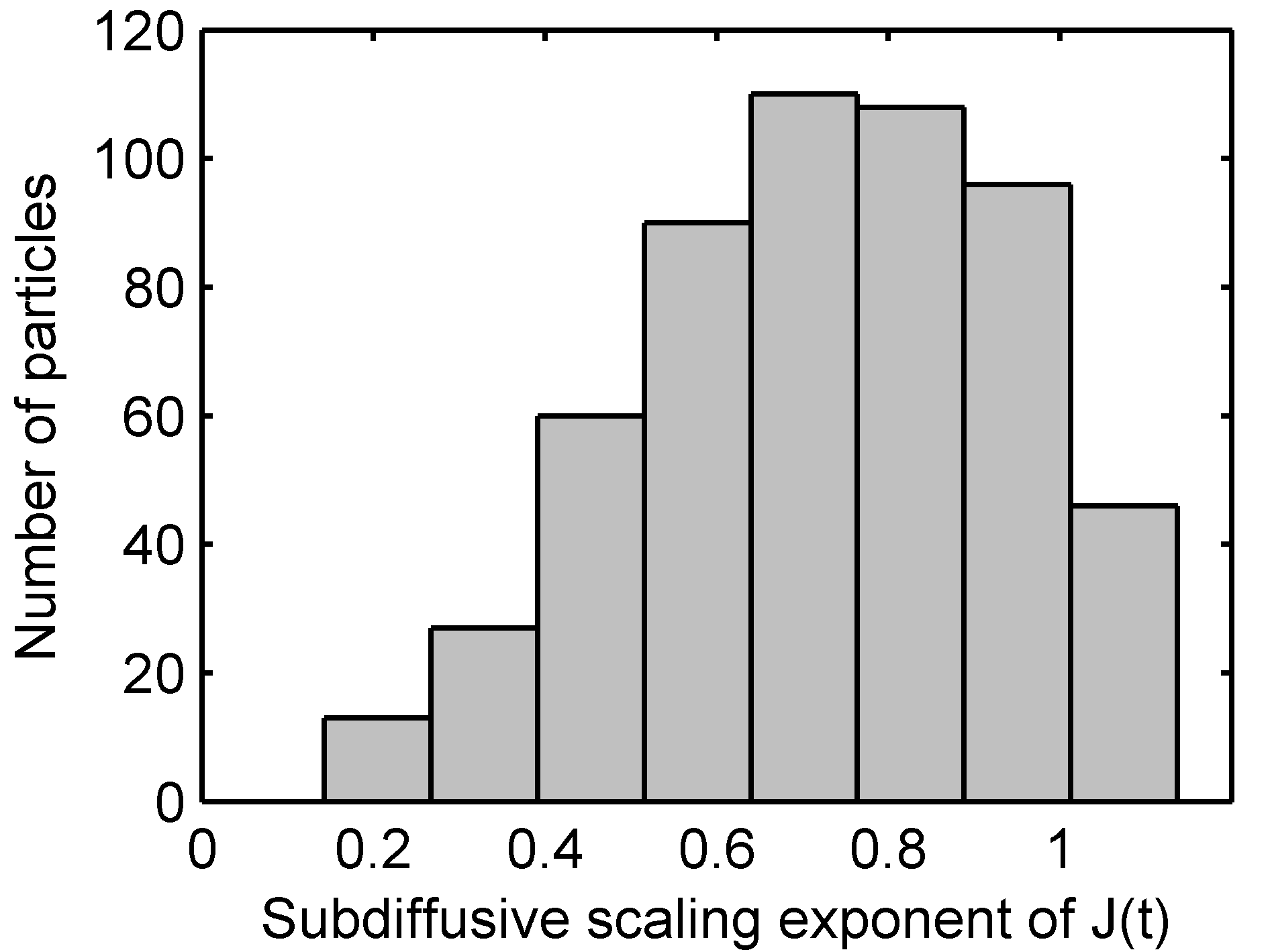}}}
\subfigure[]{\resizebox{8cm}{!}{\includegraphics{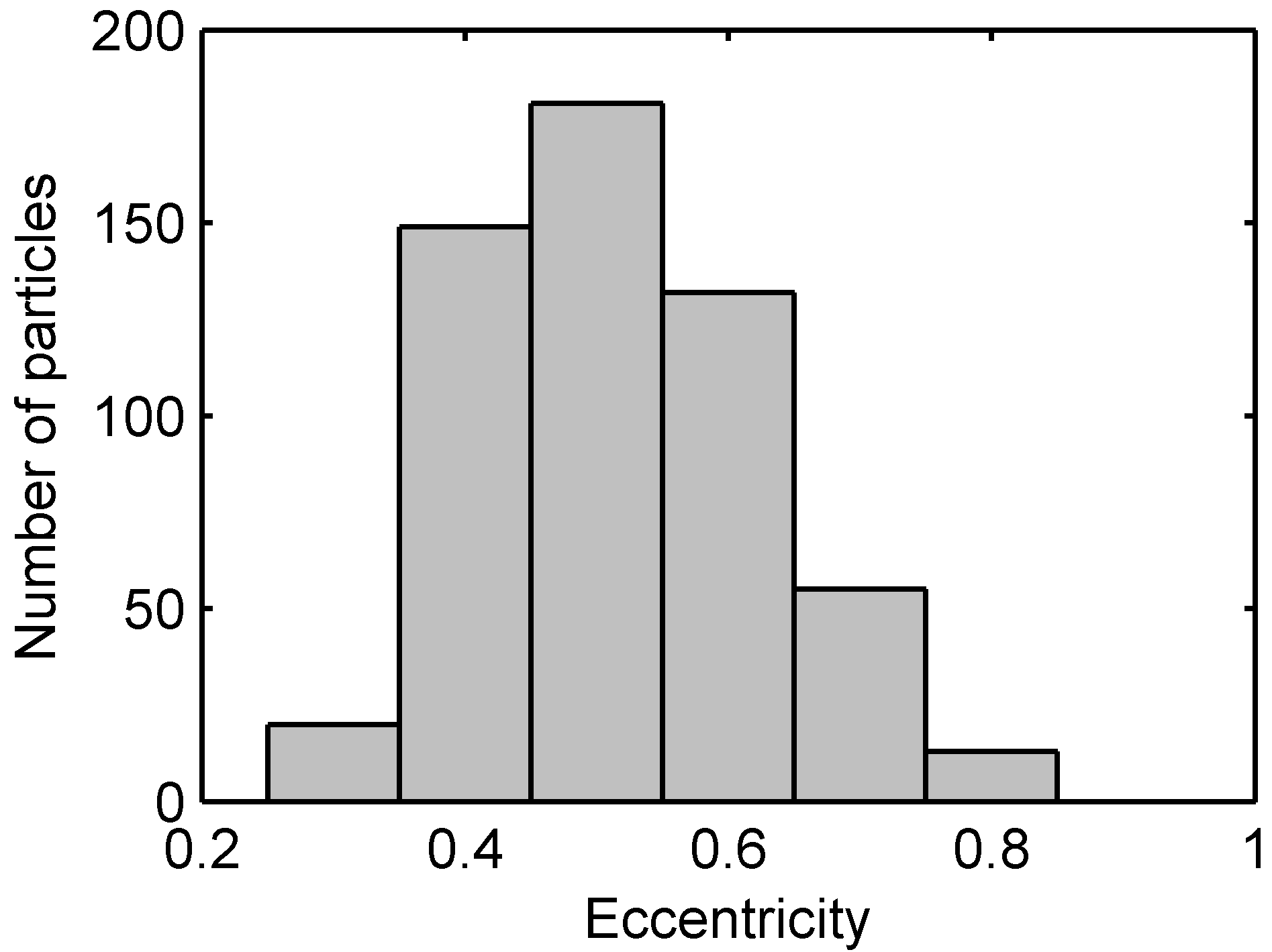}}}
\subfigure[]{\resizebox{8cm}{!}{\includegraphics{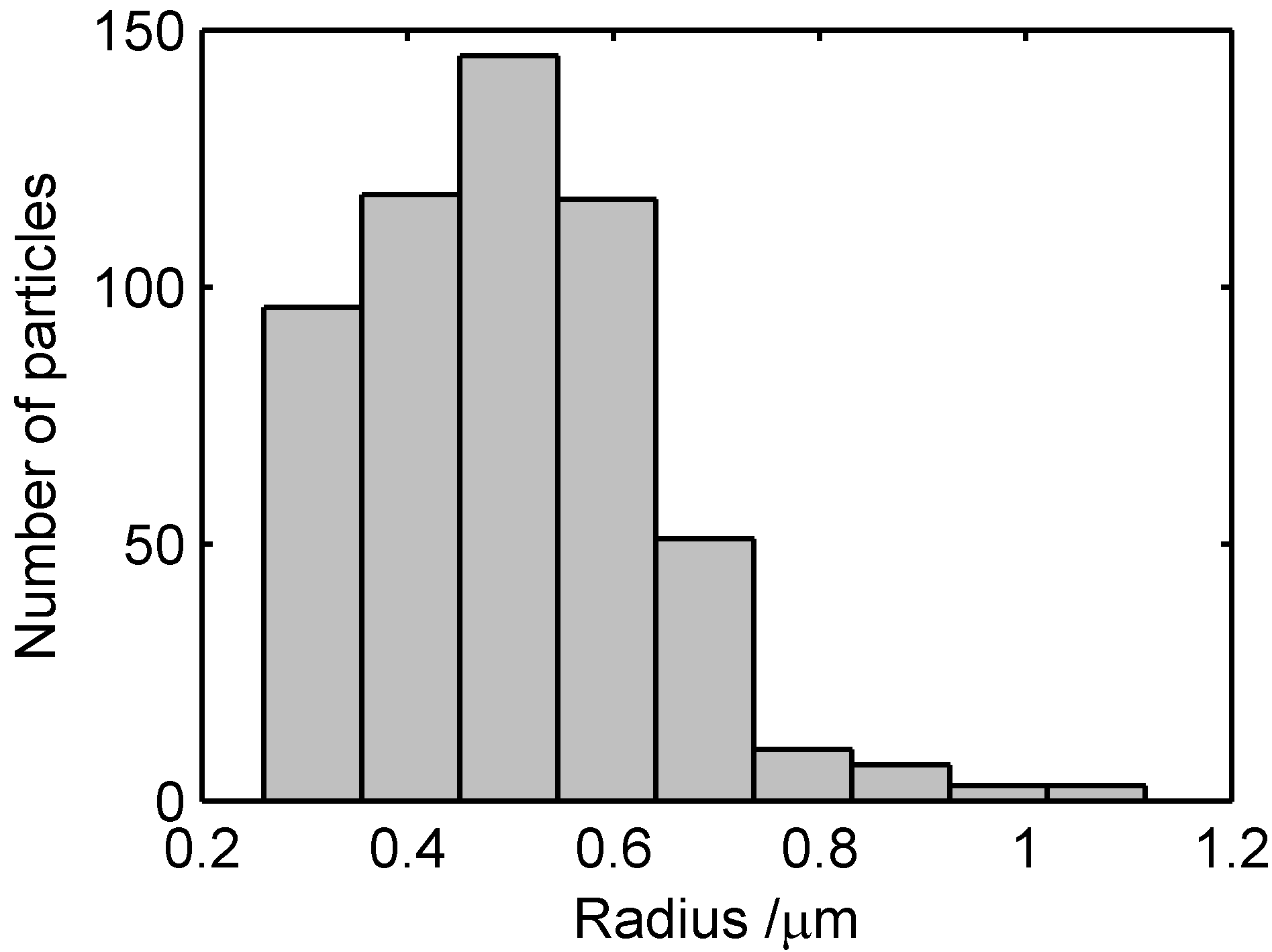}}}
\caption{(a) Histogram of scaling exponents in the subdiffusive region of $J(t)$ for each tracked particle in the small amoeba of Fig.~\ref{amoebae}(a). The histogram is peaked at 0.75, corresponding to a network of semiflexible filaments. (b-c) Histograms of radii and eccentricity of all tracked particles, calculated according to \cite{rogers2007}.}
\label{smallaprot-exponenthist}
\end{figure}

A histogram of the scaling exponents, fitted to each curve of $J(t)$ in the range $0.5\le t \le 3$~ms, is presented in Fig.~\ref{smallaprot-exponenthist}(a). The histogram is peaked at 0.75, corresponding to the tracked particles being attached or closely associated with a network of semiflexible filaments \cite{granek1997}. We also plot histograms of the radii and eccentricity of the tracked particles. The radii are peaked at 0.5~$\mu$m, and the eccentricities at 0.5. By using Eq.~\ref{eq:compliance}, we have treated all particles as spherical, therefore the eccentricity of the particles will cause an error in the analysis. Using the drag coefficients for a prolate ellipsoid \cite{perrin1934}, taking eccentricity = 0.5, we obtain a ratio of the drag coefficients, perpendicular and parallel to the principle axis, of 1.03. The error is therefore negligible compared to the heterogeneity of the cytoplasm observed above.

The same features of Fig.~\ref{smallaprot-MSD-J} were observed in all measured amoebae: i.e. a subdiffusive and a superdiffusive region, with approximately the same range of exponents and a similar range of absolute values.

\begin{figure}
\centering
\resizebox{8cm}{!}{\includegraphics{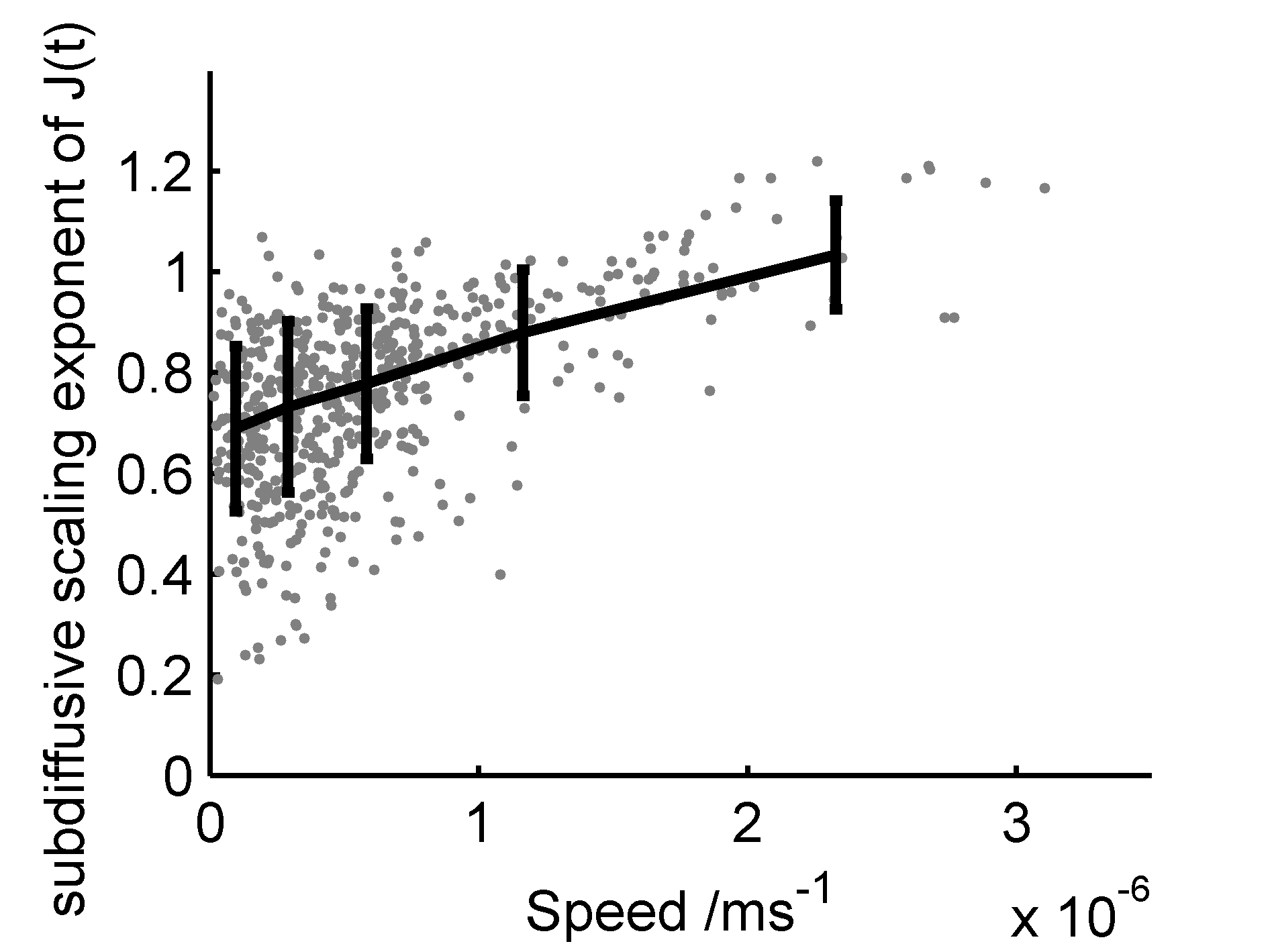}}
\caption{Scaling exponent in the subdiffusive region of $J(t)$ for each tracked particle in Fig.~\ref{amoebae}(a), plotted against particle speed (\textcolor{Gray}{$\bullet$}). The data are binned and the mean of each bin plotted as a line graph, with error bars calculated from the standard deviation of each bin. The scaling exponent shows a clear increase with increasing speed.}
\label{smallaprot-shearthinning}
\end{figure}

We investigated the dependence of the scaling exponent in the subdiffusive region on position and velocity within the cell, in order to elucidate the effect of cellular structure on rheology. Fig.~\ref{smallaprot-shearthinning} shows speed plotted against scaling exponent, calculated as above, for all tracked particles in the small amoeba above. The data points are binned, and the mean of each bin plotted, with error bars showing the standard deviation from each mean. A clear trend is apparent: the scaling exponent increases with speed, with a mean of approximately 0.7 at zero speed, and reaching a mean of 1, corresponding to a viscous liquid, at around 2~$\mu$m~s$^{-1}$. 

Note that the expected error in the subdiffusive scaling exponent due to a particle's drift velocity can be calculated as follows. For a typical particle in Fig.~\ref{smallaprot-MSD-J}(a), $\textrm{MSD} = C t^{0.75} + (Vt)^2$, where $V$ is speed in the range 0--3~$\mu$m~s$^{-1}$, and typical values of $\textrm{MSD}\approx 10^{-16}$~m$^2$ at $t=1$~ms set the parameter $C\approx 3\times 10^{-12}$. By taking the differences of the terms in this equation over the interval $0.5\le t \le 3$~ms, we calculate an error in the scaling exponent of the order $10^{-7}$, which is insignificant compared to its measured increase as a function of particle speed.

\begin{figure}
\centering
\subfigure[]{\resizebox{!}{5cm}{\includegraphics{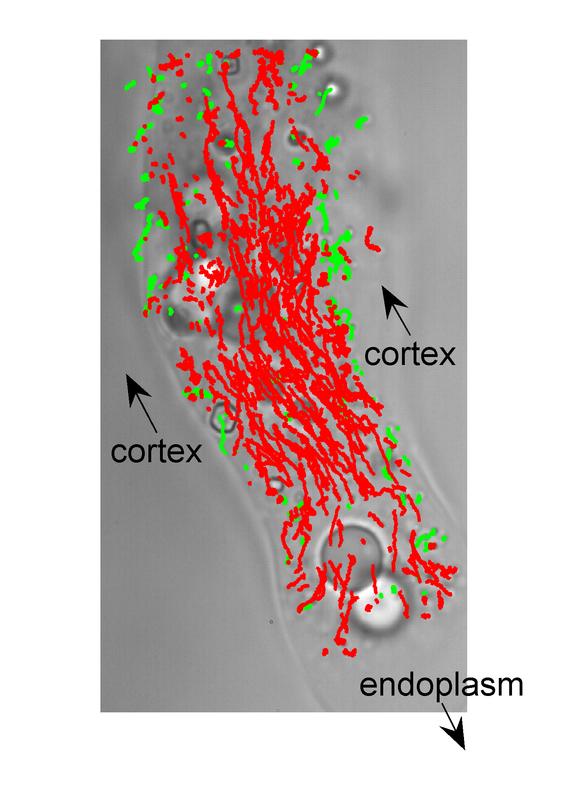}}}
\subfigure[]{\resizebox{!}{5cm}{\includegraphics{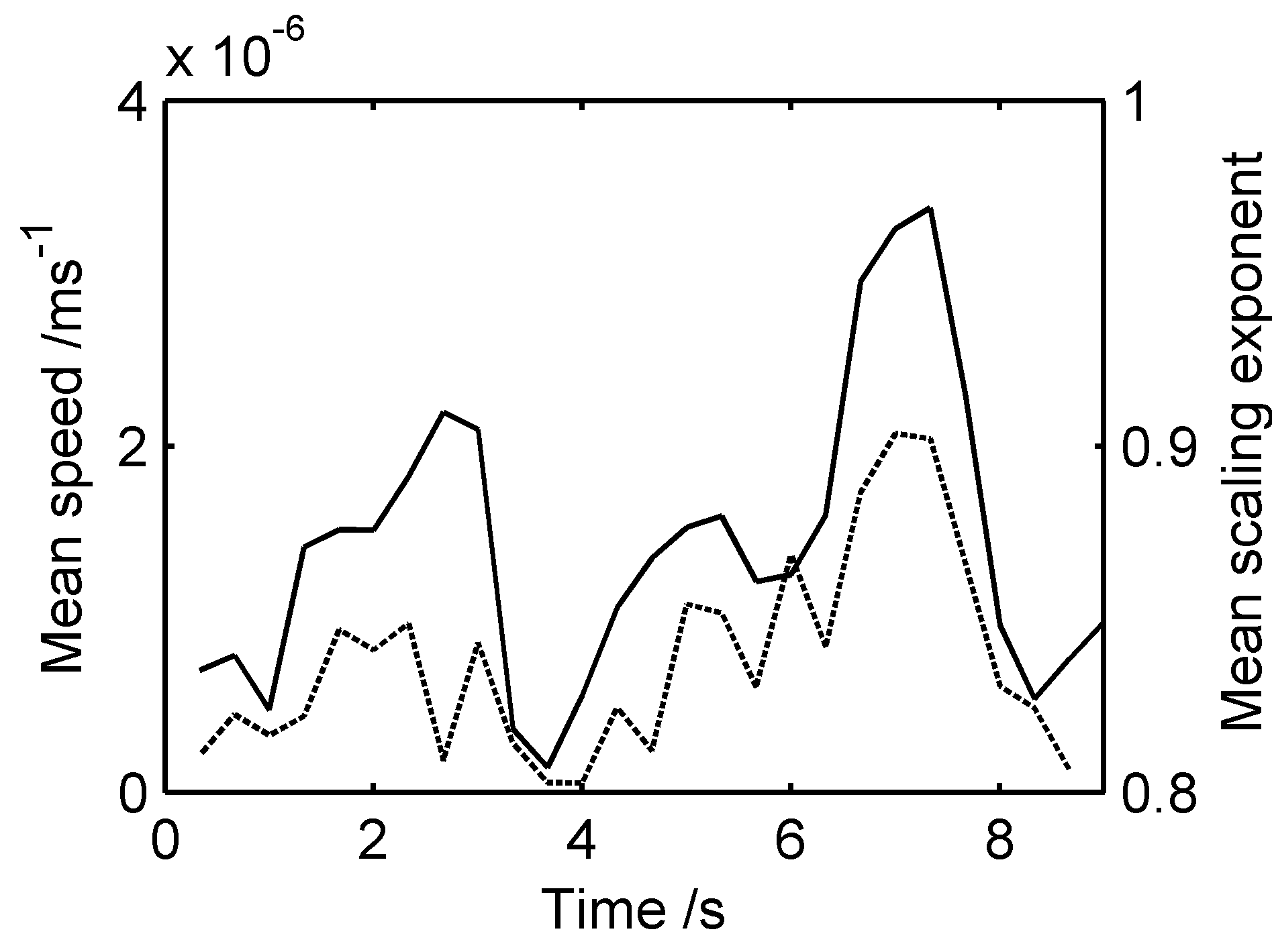}}}
\subfigure[]{\resizebox{!}{5cm}{\includegraphics{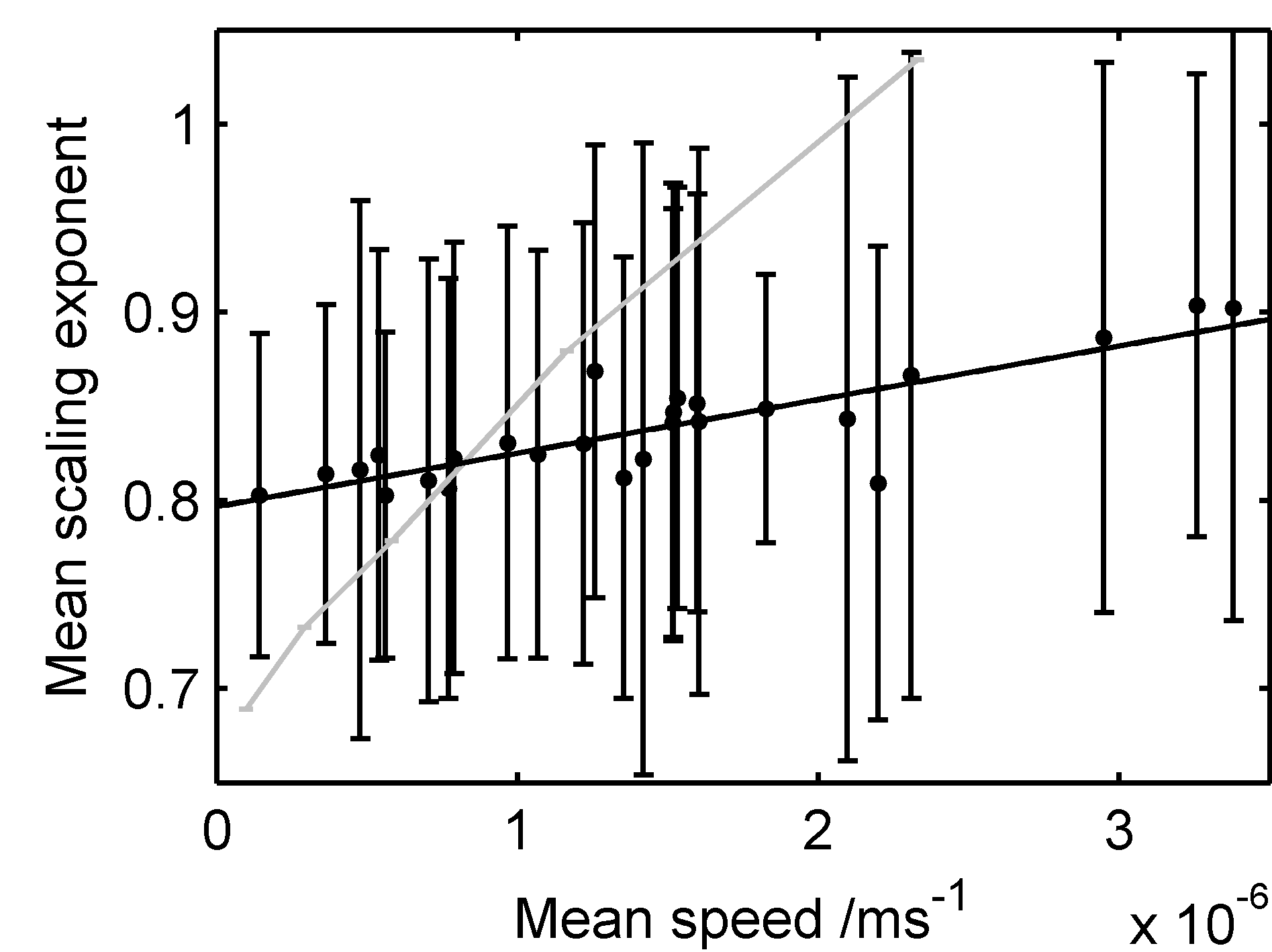}}}
\caption{(a) Endoplasmic (\textcolor{Red}{$\bullet$}) and cortical regions (\textcolor{Green}{$\bullet$}) in the lobopod of the large amoeba of Fig.~\ref{amoebae}(b), identified by the direction of particle drift with respect to the direction of the lobopod. (b) Mean particle speed of endoplasmic particles (\Flatsteel), plotted simultaneously with mean subdiffusive scaling exponent ($\dotline$), as a function of time, in steps of 1/3~s. The speed and scaling exponent rise and fall in tandem. (c) Mean particle speed and mean scaling exponent, from (b), are plotted against each other. A clear trend is shown by the linear least-squares fit (\Flatsteel): the data points follow this line much more closely than the error bars which show the standard deviation of the sample of particles at each time point. The trend is significantly shallower here than in the data for the entire small amoeba (\textcolor{Gray}{\Flatsteel}), of Fig.~\ref{smallaprot-shearthinning} above.}
\label{laprot-shearthinning}
\end{figure}

The same effect is apparent in large amoebae. Fig.~\ref{laprot-shearthinning}(a) shows the same extending lobopod of the large individual shown in Fig.~\ref{amoebae}(b). Here we highlight each particle depending on its velocity component with respect to the direction of lobopodial movement: red for positive and green for negative. Particles embedded in the endoplasm are forced by the cytoplasmic pressure in the direction of the lobopod, while particles embedded in the cortex tend to have a small velocity in the opposite direction. This well-known but counterintuitive phenomenon, known as the fountain effect \cite{allen1961a,dembo1989}, is due to the cortex of the lobopod being connected directly to the cortex of the entire cell: since the cortex is contracting everywhere, it draws the lobopodial cortex back towards the cell body. Separating the particle data by this direction of drift, we plot the speed and mean subdiffusive scaling exponent of the particles, in the endoplasm only, against time (Fig.~\ref{laprot-shearthinning}(b)). Here the data are broken up into time steps of 1000 frames (i.e.~1/3~s). The average speed of flow can be seen to pulsate rythmically, as is also apparent from the video of the lobopodial extension (see Supporting Information). The mean subdiffusive scaling exponent, calculated as above, rises and falls in tandem with the speed, and they are plotted against each other in Fig.~\ref{laprot-shearthinning}(c). Here, a clear trend is shown by the linear least-squares fit (\Flatsteel): the data points follow this line much more closely than the plotted error bars, which show the standard deviation of the sample of particles at each time point. Indeed this standard deviation is an overestimate of the statistical error in this plot, since it is due to the inhomogeneity in the local environment of each particle, which does not necessarily change between successive time points. In the lobopod endoplasm only, the trend is significantly shallower than in the data for the entire small amoeba, of Fig.~\ref{smallaprot-shearthinning} above, which is superposed here (\textcolor{Gray}{\Flatsteel}). These data show an analogous phenomenon to shear thinning, and are further discussed below, along with the differences between the measurements.

\begin{figure}
\centering
\subfigure[]{\resizebox{8cm}{!}{\includegraphics{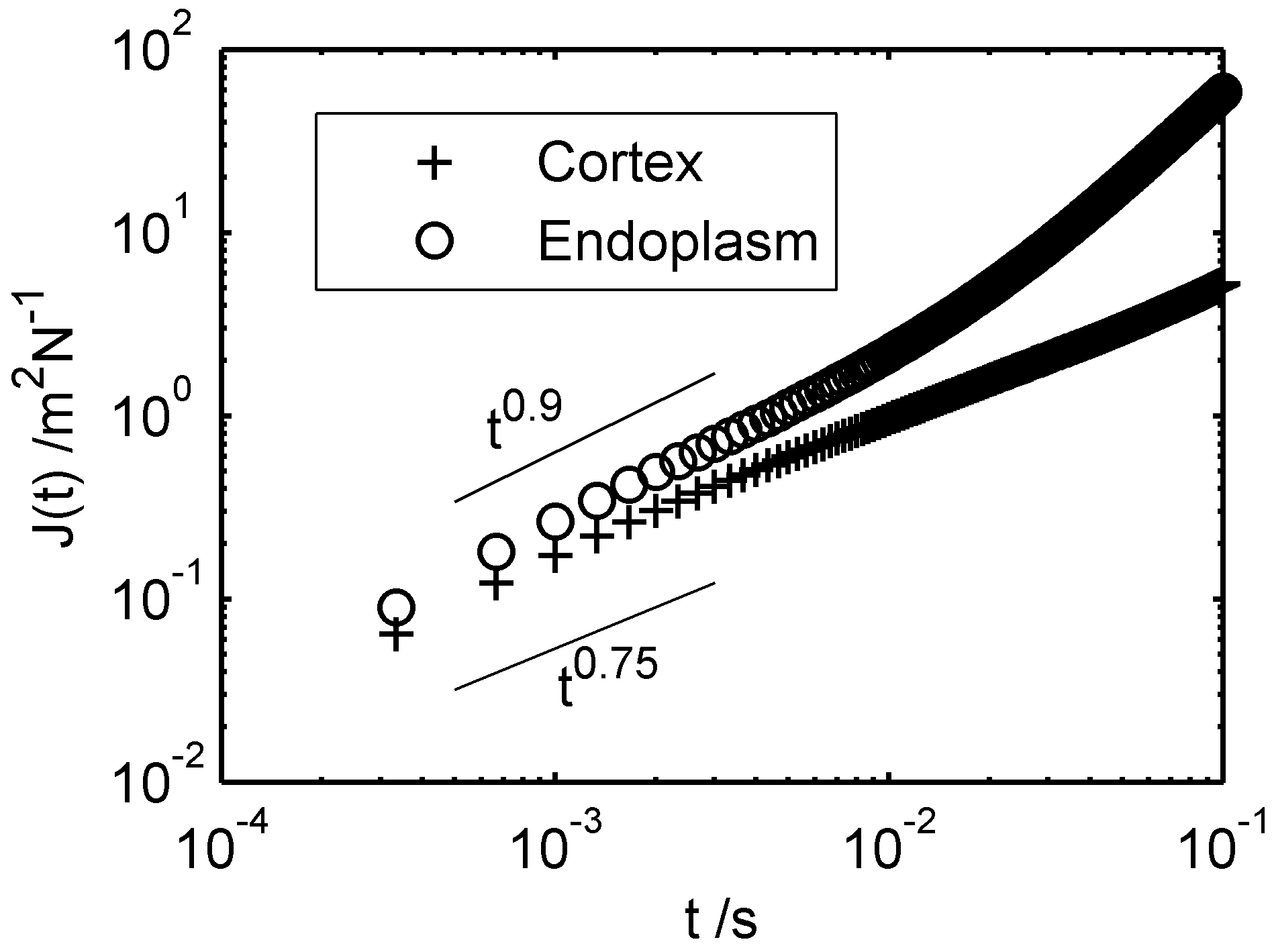}}}
\subfigure[]{\resizebox{8cm}{!}{\includegraphics{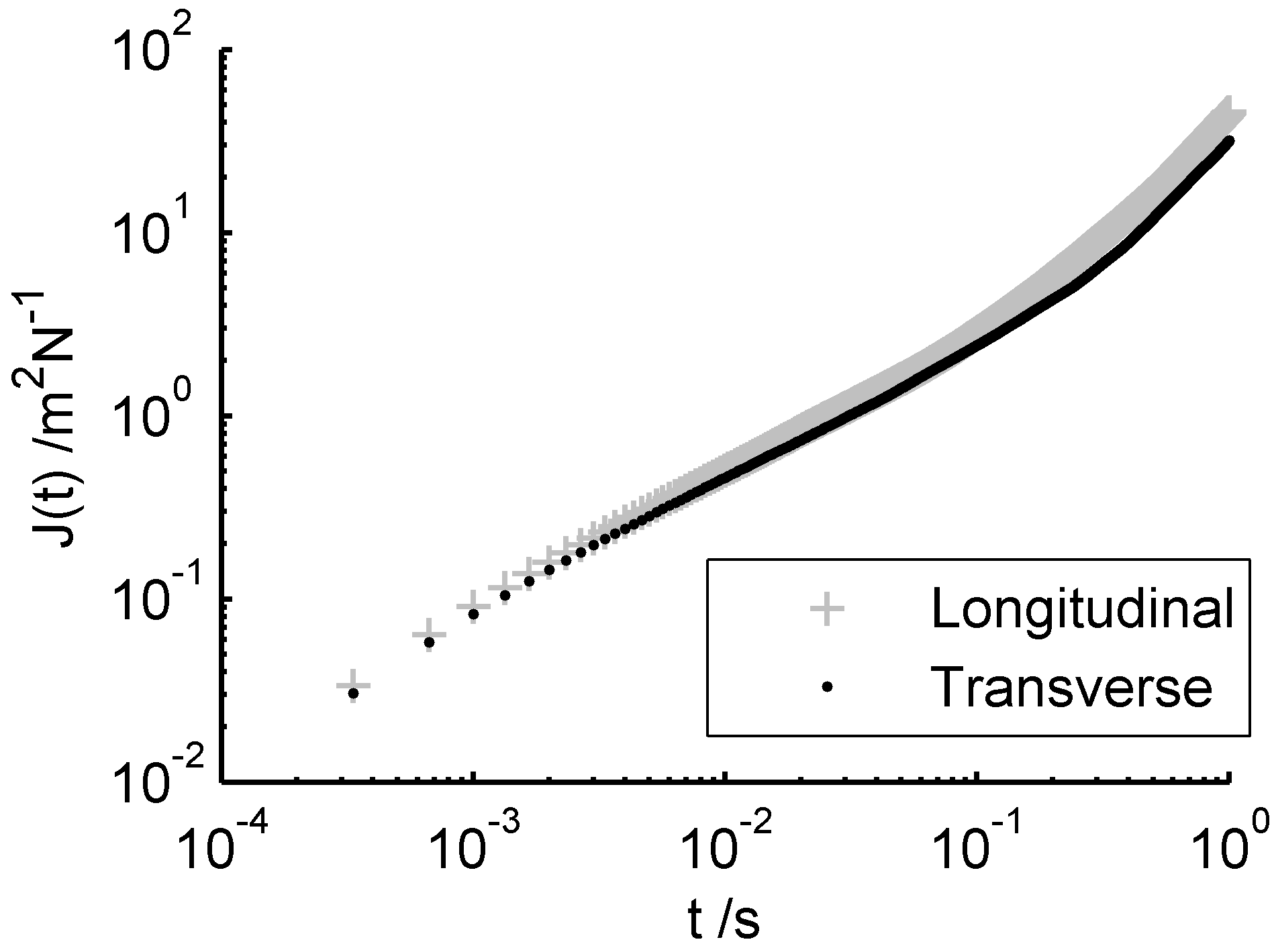}}}
\caption{(a) Compliance within the cortex and endoplasm of the lobopod shown in Fig.~\ref{laprot-shearthinning}(a). There is a small difference in the subdiffusive regime: the mean scaling exponents are close to 0.75 and 0.9 for cortex and endoplasm respectively. (b) Rheological anisotropy in the cortex shown as effective compliance in the longitudinal and transverse directions with respect to the lobopod. There is a negligible difference: the rheology of the cortex is effectively isotropic.}
\label{endocortex}
\end{figure}

The effect of structure on rheology within the lobopod can be further investigated. In Fig.~\ref{endocortex}(a) we plot the compliance within the cortex and endoplasm respectively, averaged over all particles separated as shown in Fig.~\ref{laprot-shearthinning}(a). Looking only at the subdiffusive regime, we see that there is a small difference: the mean scaling exponents are close to 0.75 and 0.9 for cortex and endoplasm respectively. Indeed the difference in the absolute value of $J(t)$ is fairly small: the endoplasm is 54\% more compliant than the cortex on a time scale of 1~ms, and the compliances of both sections appears as if they will converge to the same value if extrapolated to $\sim 10^{-4}$~s. We can fit endoplasmic compliance with a linear function of time, since its scaling exponent of the endoplasm is so close to 1. Fitting $J(t)$ in the range $0.5\le t \le 3$~ms, we obtain $J(t) \approx 220 \times t$~Pa$^{-1}$s$^{-1}$. Thus the effective viscosity is given by $\eta = t/J(t)= 4.5$~mPa~s. The rheology within a lobopod section is further explored below in a different individual. 

Besides rheological inhomogeneity, there may be rheological anisotropy within the cytoplasm. In particular, we may expect the actin filaments within the cortex to have an anisotropic arrangement, such as a preferred orientation, parallel or perpendicular to the lobopod. Any such anisotropy in cortical structure would surely be reflected in the rheology. Following \cite{hasnain2006}, we analyse rheological anisotropy by treating the MSD as a tensor: $\langle (r_i(T+t)-r_i(T)) (r_j(T+t)-r_j(T))  \rangle$, where $i,j=\{x,y\}$. By rotating the coordinate axes to the directions parallel and perpendicular to the lobopod, and treating each direction independently, we may use Eq.~\ref{eq:compliance} to obtain the effective compliance in each direction. Fig.~\ref{endocortex}(b) shows the result: there is negligible difference between the compliance in each direction; indeed the difference is very much smaller than the standard deviation of compliances between different particles. Therefore, the cortex may be considered rheologically isotropic. The same result was obtained on two other measured lobopodia.

\begin{figure}
\centering
\resizebox{8cm}{!}{\includegraphics{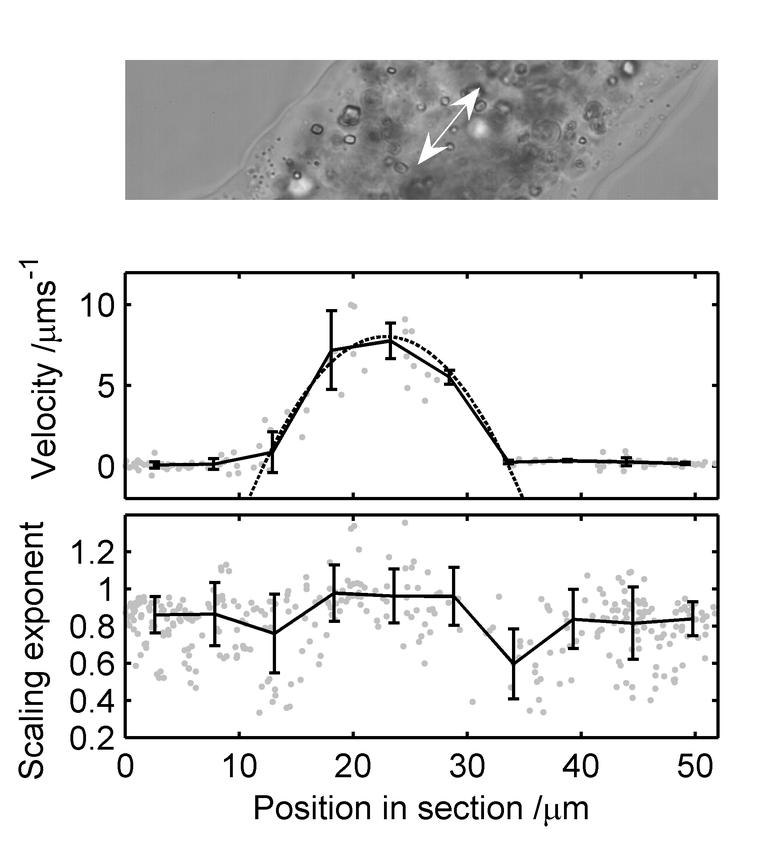}}
\caption{The mid-section of an \emph{A.~proteus} lobopod is displayed (top), showing the direction of flow. From tracked particles in this image, we calculate a velocity profile (centre) and a scaling exponent profile (bottom). Individual data points (\textcolor{Gray}{$\cdot$}) are plotted on each graph, with the rebinned and averaged data plotted with error bars, showing each standard deviation from the mean (\Flatsteel). The velocity of the endoplasmic layer is fitted well with a parabola ($\dotline$), which corresponds to Poiseuille flow in a Newtonian fluid. The scaling exponent profile shows two regions of scaling exponent $\approx$~0.8 and 1.0, respectively corresponding to the cortex and endoplasm as identified from the velocity profile. The boundary between the layers seems abrupt; there is no gradual change in the mean-scaling-exponent, as far as the local inhomogeneity in the data permits us to discern.}
\label{velprofile}
\end{figure}

The velocity profile and subdiffusive scaling exponent profile are plotted simultaneously for the mid-section of another \emph{A.~proteus} lobopod in Fig.~\ref{velprofile}(top). The data were taken from an 8-s movie captured at a framerate of 3000~Hz, during which the flow rate varied continually. The velocity profile was calculated from all tracked particles during the last 2~s of the movie, when the flow rate was approximately constant. We plot the component of the velocity parallel to the lobopod (Fig.~\ref{velprofile}(centre)): here we can see the nearly sessile cortical layer, approximately 15~$\mu$m thick, and the flowing endoplasm, with a diameter of around 20~$\mu$m. The data points are rebinned, and the mean and standard deviation in each bin are plotted as a solid line with error bars (\Flatsteel). The velocity of the endoplasm shows good correspondence to a parabolic fit ($\dotline$), which would be indicative of a Poiseuille flow in a Newtonian fluid in contrast to the plug flow (flattened parabola) exhibited by strongly elastic suspensions \cite{Isa2007}. The scaling exponent profile is calculated as above, from all tracked particles in the entire movie (Fig.~\ref{velprofile}(bottom)). Again, the data are rebinned and the means and standard deviations plotted as a solid line with error bars. This profile shows two regions of scaling exponent $\approx$~0.8 and 1.0 respectively: these correspond to the cortex and endoplasm as identified from the velocity profile. Although the statistics of each bin limit our resolution to $\sim 5$~$\mu$m in analysing the spatial variation of the rheology, it appears here that the boundary between cortex and endoplasm is fairly abrupt: no gradual change in the mean-scaling-exponent is apparent, and the boundary itself has an upper thickness limit of 5~$\mu$m.

\begin{figure}
\centering
\resizebox{8cm}{!}{\includegraphics{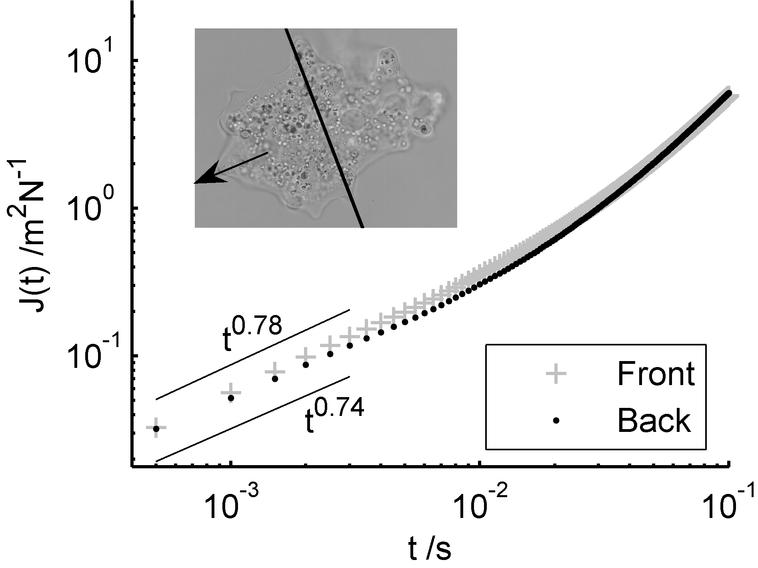}}
\caption{The particle tracks from the individual of Fig.~\ref{amoebae}(a) are bisected into front (extending) and back (receding) halves. Thence the mean compliance in each region is plotted. There is a small difference in average rheology between the extending and receding regions: in this amoeba, the fitted scaling exponents were 0.78 and 0.74 in the front and back regions respectively, but this difference varied between different amoebae (not shown).}
\label{frontandback}
\end{figure}

We looked for differences in the rheology of the cortex and endoplasm in extending and retracting parts of \emph{A.~proteus}. The small amoebae could be completely observed in a single field of view under the microscope, so that it was simple to bisect the images into front (extending) and back (receding) halves, and measure the microrheology of each. Fig.~\ref{frontandback}(inset) shows the bisection of the data from the individual of Fig.~\ref{amoebae}(a), thence the mean compliance in each region is plotted. There is a very small difference in average rheology between the extending and receding regions---far less than the standard deviation from the mean (not shown). In this amoeba, the fitted scaling exponents were 0.78 and 0.74 in the front and back regions respectively, but this small difference was within the error bars on repeat measurements with different amoebae (not shown).

\begin{figure}
\centering
\resizebox{8cm}{!}{\includegraphics{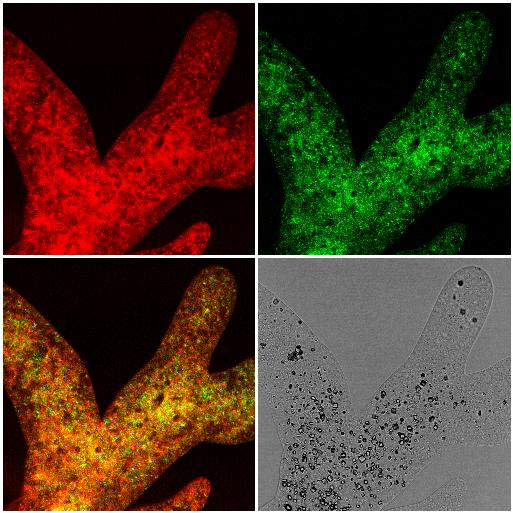}}
\caption{Large individual of \emph{A.~proteus} fixed in formaldehyde and stained for microtubules (green) and F-actin (red). The fluorescence images are overlaid (bottom-left), and a corresponding bright field image is shown (bottom-right). F-actin is distributed throughout the cell, while tubulin is predominantly found deeper within the cell.}
\label{fluor}
\end{figure}

Samples of \emph{A.~proteus} were fixed and stained for microtubules and filamentous actin as described above. Fig.~\ref{fluor} shows representative images. Phalloidin-stained actin is found throughout the cell, while fluorescently-labelled taxol seems to have stained particulate matter rather than extended microtubules. This confirms previous observations of the lack of microtubules in the cytoplasm \cite{taylor1979}. The observed particular matter is likely to be short microtubules and microtubule-organising centres \cite{gromov1985}. No layer of high actin concentration is observed at the cell periphery---indeed the observation of its absence in immobilised cells has been noted previously \cite{stockem1983}. Only in live cell fluorescence images does a layer of high actin concentration develop transiently near the cell periphery and continually change during cell movement \cite{stockem1983,hoffmann1984}. It seems that filamentous actin is present throughout the cell, and concentrated layers of actin only appear during cortical contractions before disintegrating again.

\section{Discussion}

Although the study of amoeboid motion in general and \emph{A.~proteus} in particular are very old topics, our application of microrheology to them has yielded novel results. Firstly we identified a power law viscoelastic regime in $J(t)$ at short time scales, similar to that observed in cultured animal cells and reconstituted actin networks. Although there was a large variability in the absolute value of $J(t)$ due to inhomogeneity of the cytoplasm and presumably heterogeneity of interactions between each particle and the cytoskeleton, we found that the exponent of the power law regime of $J(t)$ gave illuminating results. The mean of the distribution of exponents was near 3/4 in all cases where the cytoplasm was at rest. A similar result has been obtained several times in reconstituted actin networks \cite{amblard1996,gisler1999,koenderink2006}, but rarely in cultured mammalian cells, where very different exponents are reported \cite{Balland2006}, although the famous exponent of 3/4 has been found by some authors \cite{Hoffman2006,Mizuno2007}: we feel that these discrepancies highlight the difficulties of accurate intracellular microrheology. However, the exponent of 3/4 can be observed more clearly in our data than any other published data of which we are aware. This success is due to the size of \emph{A.~proteus}, the large number of trackable particles it contains, and the use of a fast digital camera.

We then examined the effect of flow on the rheology of the cytoplasm, finding in each case that the $J(t)$ scaling exponent increased with flow speed: i.e.~the cytoplasm became more fluid-like. In the case of pulsating flow in the large amoeba lobopod, the shear rate can be estimated by assuming Poiseuille flow in a cylinder of diameter 10~$\mu$m, the approximate diameter of the endoplasm in Fig.~\ref{amoebae}(b). Thus the range of speeds 0--3.4~$\mu$m~s$^{-1}$ maps onto the range of shear rates 0--0.7~s$^{-1}$. One obvious reason that the measurements of shear-thinning in figure \ref{laprot-shearthinning}(c) do not agree with each other, is that the geometry of the flow is different in the different amoebae, although unfortunately we cannot estimate the shear rates in the small amoeba because we lack data on the velocity of the flow as a function of height above the substrate. In the endoplasm of the large amoeba, the flow rate rose and fell on a time scale of $\sim$1~s, and the scaling exponent followed it without a measurable delay. On that time scale, the change in rheology could be due to either passive or active remodeling of the cytoskeleton \cite{Mizuno2007}. We should additionally expect that the active remodeling of the cytoskeleton occurs on a range of longer time scales; for example, the active cycling of material between the cortex and endoplasm in the course of cell migration occurs on a time scale of minutes \cite{stockem1983}.

These measurements join very few previous reports of shear-thinning in living cytoplasm, which measured the apparent viscosity under forces applied with magnetic tweezers \cite{Valberg1987,Marion2005} and micropipettes \cite{Tsai1993}. However, our measurements of the effect of flow on the rheological scaling exponent are new, and reveal information that could be compared with theories of viscoelasticity of semiflexible fibre networks. At present, no such theories exist of how the rheology of these networks changes with deformation. But heuristic models of the shear-thinning response of cells are already in use, as they are necessary to explain the measured deformations of whole cells under applied forces, such as during micropipette aspiration \cite{Lim2006}.

Comparing the rheology of cortex and endoplasm, we found that the absolute values of compliance were similar in both, and may be expected to converge to the same value at time scales $\lesssim 10^{-4}$~s. (Fig.~\ref{endocortex}(a)). The similarity in compliance between the two regions is perhaps unsurprising given the fairly uniform distribution of filamentous actin throughout the cell (Fig.~\ref{fluor}). The rheology of the endoplasm was found to be close to a Newtonian liquid with a viscosity of 4.5~mPa~s. Although reported values of cytoplasmic viscosity are known to vary by six orders of magnitude (quoted in \cite{Valberg1987,Yamada2000}), our value is similar to several measurements at the lower end of the scale. It is not surprising that these techniques accessing different parts of the cell machinery at different time scales, length scales and deformation rates have yielded such different values of effective viscosity. But our measurement is notable in that we know it describes the flowing endoplasm only, whose rheology we have found to be approximately Newtonian. If we use this value to obtain the Reynolds number and P\'{e}clet number of the endoplasmic flow, we obtain $\textrm{Re}=\rho VR/\eta \approx 10^{-5}$ and $\textrm{Pe}=LV/D \approx 10^3$, taking the radius of the channel as $R=10~\mu$m, the velocity as $V=5~\mu$m~s$^{-1}$, the density $\rho$ of water, and the typical diffusivity of a tracked particle as $D=k_b T/ 6\pi a \eta$ with $a\approx 1~\mu$m. The Reynolds number is low---there is no turbulence, and the P\'{e}clet number is high---convection dominates the particle motion over the length scale $R$. A typical pressure drop can be estimated along a lobopod of length $L=100~\mu$m: $\Delta P= 4 \eta V L/R^2 \approx 0.1$~Pa, using the equation for Poiseuille flow. We note that the pressure difference within the endoplasm is very much smaller than the pressure differences of 10$^2$--10$^3$~Pa required to stall the amoeba's motion \cite{tasaki1964,grebecka1980,yanai1996}: although large stresses are generated within the cortex, the endoplasm transmits the pressure hydrostatically.

In Fig.~\ref{endocortex}(b), we were surprised to find that the rheology of the cortex is effectively isotropic, suggesting that the actin filaments in the cortical cytoskeleton are arranged isotropically, rather than with orientational order with respect to the cell geometry. This novel result provides reassuring evidence that theories and measurements in vitro of model isotropic actin networks correspond to the system in vivo, at least in the case of \emph{A.~proteus}.

The velocity profile and scaling exponent profile in the lobopod section (Fig.~\ref{velprofile}) show that the streaming amoeba endoplasm flows like Newtonian fluid, with a parabolic velocity profile and a scaling exponent of 1 throughout the endoplasm. The boundary between the endoplasm and cortex seems intriguingly abrupt in both profiles: we are not aware that the physics of this boundary has been studied. The scaling exponent was near 0.75 within the cortex as found previously.

Finally, we saw that the average rheology of a small amoeba did not differ significantly when bisected into front and back halves. In both halves, the average was dominated by the particles embedded in the cortex, shown by the exponent of $\approx 0.75$. Although the back half of the cortex must have a larger tension than the front half in order to produce the observed motion, there is no significant difference in the rheology of the two halves. This result suggests that the difference in tension is not caused by a difference in the structure of the cytoskeletal network which would surely affect the microrheology. Rather it is likely to be caused simply by the difference between the thickness of the cortex in the two halves. We note that the result contrasts with the microrheology of migrating animal fibroblasts: Kole et al.~\cite{kole2005} found that the leading part (the lamellipodium) of 3T3 cells is significantly less compliant than the trailing part, which contains the nucleus and most of the cytoplasm. Following them, we found that the lamellipodia of 3T3 and HeLa cells have lower rheological scaling exponents than the perinuclear region (unpublished data). It is not clear if these differences within animal cells are due to the structure of the cytoskeleton or the confinement of the particle and tight association to the substrate within a thin lamellipodium.

We may compare the measurements of compliance in \emph{A.~proteus} with previous PTM studies on mammalian cells and F-actin solutions. Measurements on COS7 cells \cite{Yamada2000} and 3T3 \cite{kole2005} yielded compliance in the range 0.01--0.1~m$^2$N$^{-1}$ on a time scale of 0.1~s. In the different individuals of \emph{A.~proteus} that we have sampled, we have found compliances in the range 1--10~m$^2$N$^{-1}$ on a time scale of 0.1~s, similar to measurements of F-actin solutions, at physiological concentrations \cite{Yamada2000,Gardel2003}. The cytoplasm of \emph{A.~proteus} is therefore less stiff than these typical mammalian cell lines, by a factor of $\sim$~100.

\section{Conclusion and outlook}

Our application of PTM to crawling \emph{A.~proteus} has yielded novel results on the structure and rheology of the cytoskeleton in living cells. In particular, we have seen variations (or lack of variations) of the rheological scaling exponent with flow rate and position or orientation within the cortex, endoplasm and extending or contracting parts of the cell. \emph{A.~proteus} has always been one of the major organisms for studying amoeboid motility. Its resemblance to mammalian cells in many types of motion, such as cortical oscillations, make our results on its microrheology relevant to the wider field of cell motility. The endoplasm of \emph{A. proteus} behaves as Newtonian fluid at fast flow rates, and has a velocity profile close to Poiseuille flow. The compliance of the cortex displays a clear $t^{3/4}$ exponent, which is thought to be indicative of the transverse bending fluctuations of actin filaments.

\section*{Acknowledgements}
This project was funded by the UK EPSRC under grant no.~EP/E013988/1. Thanks to Marcus Jahnel, Xiubo Zhao, Daniel Sate and Paul Coffey for many fruitful discussions.

\bibliographystyle{unsrt}
\bibliography{microrheoaprot}

\end{document}